\documentclass[twocolumn, showpacs, prl]{revtex4}

\usepackage{graphicx}% Include figure files

\usepackage{dcolumn}% Align table columns on decimal point

\usepackage{bm}% bold math

\begin{document}

\title{Temperature dependence of piezoresistance of composite Fermions with a valley degree of freedom}

\date{\today}

\author{T.\ Gokmen}

\author{Medini\ Padmanabhan}

\author{M.\ Shayegan}

\address{Department of Electrical Engineering, Princeton
University, Princeton, NJ 08544}

\begin{abstract}

We report transport measurements of composite Fermions at filling
factor $\nu=3/2$ in AlAs quantum wells as a function of strain and
temperature. In this system the composite Fermions possess a valley
degree of freedom and show piezoresistance qualitatively very
similar to electrons. The temperature dependence of the resistance ($R$) of composite Fermions shows a metallic behavior ($dR/dT > 0$) for small values of valley
polarization but turns insulating ($dR/dT < 0$) as they are driven
to full valley polarization. The results highlight the importance of
discrete degrees of freedom in the transport properties of composite Fermions and the similarity between composite Fermions and electrons.

\end{abstract}

%\begin{keyword}

%A. Quantum wells; D. Fractional quantum Hall effect; D. Electronic transport; E. Strain, high pressure

%\end{keyword}

\maketitle

Since the discovery of the fractional quantum Hall effect (FQHE)
\cite{TsuiPRL82}, a great deal of research has been devoted to
understand the ground state of a two-dimensional electron system
(2DES) at high magnetic fields. Although Laughlin's original wave
function successfully explained the first observed FQHE state at
filling factor $\nu=1/3$ \cite{LaughlinPRL83}, it is the composite
Fermion theory \cite{JainPRL89,KalmeyerPRB92,HalperinPRB93,CFbook}
that has unified the origin of nearly all the fractional states.
Composite Fermions (CFs) are formed by attachment of an even number
of magnetic flux quanta to each electron. At exact half-fillings the
attached flux cancels out the external magnetic field and the CFs
feel zero \textit{effective} magnetic field. They are therefore
expected to have Fermi liquid properties and, in particular, form a
Fermi sea \cite{HalperinPRB93,CFbook}; this has indeed been verified
in numerous experiments \cite{WillettPRL93,KangPRL93,GoldmanPRL94}.

Although the flux attachment cancels the external magnetic field at
half-fillings, any spatial inhomogeneity in the density of electrons
(because of the random impurity potential) results in a random,
non-zero effective magnetic field that is seen by CFs. Such a random
field is expected to suppress the weak localization effect and give
rise to a metallic ground state for CFs in a low-disorder 2DES
\cite{KalmeyerPRB92}. With increasing disorder, the CF system can be
driven through a metal-insulator transition (MIT)
\cite{KalmeyerPRB92}. Indeed, for CFs at $\nu=1/2$ a metallic
temperature dependence at high densities and a disorder-induced MIT
(via lowering the density) was experimentally demonstrated
\cite{LiangSSC1997}.

Here we report piezoresistance measurements for CFs at $\nu=3/2$ in
an AlAs quantum well 2DES. In this system, the CFs possess a valley
degree of freedom \cite{BishopPRL07}, and their valley occupation
can be controlled via the application of in-plane strain. Our
piezoresistance traces at $\nu=3/2$ show an increase in the
resistance of CFs and a clear "kink" as the CFs make a two-valley to
single-valley transition. This is qualitatively similar to the
piezoresistance of electrons at zero magnetic field. The temperature
dependence of the piezoresistance reveals that, like their electron
counterparts \cite{GunawanNature}, increasing valley polarization changes the sign of the temperature dependence of resistance of CFs signaling the importance of the discrete degrees of freedom in the transport properties of CFs.

We performed experiments on a 2DES confined to a 15 nm thick layer
of AlAs, and modulation-doped with Si. Our sample was grown by
molecular beam epitaxy on a (001) GaAs substrate. The electrons in
this sample occupy two in-plane valleys with elliptical Fermi
contours as shown in Fig. 1(b) \cite{ShayeganPSS2006}, each centered
at an X point of the Brillouin zone, and with an anisotropic
effective mass (longitudinal mass $m_{l}=1.05$ and transverse mass
$m_{t}=0.205$, in units of free electron mass). We refer to these
valleys by the orientation of their major axis, [100] and [010]. To
vary the occupations of the two valleys we glue our samples to a
piezoelectric actuator (piezo), and apply voltage bias to the piezo
to stretch the sample in one direction and compress it in the
perpendicular direction
\cite{BishopPRL07, GunawanNature, ShayeganPSS2006, GunawanPRL2006, ShkolnikovAPL2004}.
This results in a symmetry breaking strain $\epsilon =\epsilon
_{[100]} - \epsilon _{[010]}$, where $\epsilon_{[100]}$ and
$\epsilon_{[010]}$ are the strain values along the [100] and [010]
directions. For $\epsilon > 0$ electrons are transferred from the
[100] valley to the [010] valley and vice-versa for $\epsilon < 0$;
in either case the total density remains fixed with strain. The
resulting valley splitting energy is given by $E_V = \epsilon E_2$
where $E_2$ is the deformation potential which in AlAs has a band
value of 5.8 eV \cite{ShayeganPSS2006}. We use a metal-foil strain
gauge glued to the opposite face of the piezo to measure the applied
strain \cite{ShayeganPSS2006}. The lithographically defined Hall-bar
mesa is aligned along the [110] direction to pass current at
$45^{\circ}$ with respect to the major axes of the valleys so that
the antisymmetric piezoresistance due to mass anisotropy
\cite{ShkolnikovAPL2004} is minimized. We used a top gate to vary
the electron density ($n$). All measurements were done in a dilution
refrigerator with a base temperature ($T$) of 20 mK and an 18 T
superconducting magnet.

In Fig. 1(a) we show resistance ($R$) vs. magnetic field ($B$)
data at $n=5.47 \times 10^{11}$ cm$^{-2}$ for $\epsilon = 0$
(balanced valleys). In addition to the integer quantum Hall states
at low $B$, well-developed FQHE states around $\nu=3/2$ such as
$\nu=$ 5/3, 4/3, 8/5 and 7/5 (and also in the second Landau level at
$\nu=8/3$ and 7/3) can be seen. After determining the density and
the position of $\nu=3/2$ from magnetoresistance data, we take
piezoresistance traces at $B=0$ and at $\nu=3/2$ as shown in Figs.
1(d) and (e). Each trace in these figures is normalized to the value
of resistance at $\epsilon = 0$ and shifted by 0.3 units vertically
for clarity. In the remainder of the paper we first discuss the
piezoresistance data taken at $B=0$ and then come back to $\nu=3/2$.
Finally, we present data on the temperature dependence of the
piezoresistance.

The Fermi contour of the electrons at zero strain, and the $B=0$
piezoresistance data for a range of densities are shown in Figs.
1(b) and (d), respectively. The piezoresistance data show two noteworthy features:
First, the resistance increases as strain is swept away from zero.
Second, at high values of strain the resistance shows a kink
following which it changes very slowly. (Traces
taken at high densities do not show the kink because of our limited
strain range.) The kink positions in the piezoresistance traces mark
the onset of full valley polarization of electrons, as documented
previously \cite{GunawanNature}. The increase in resistance is
because of the transfer of electrons between the valleys and can be
understood reasonably well by incorporating the anisotropic
effective mass of electrons \cite{DordaPRB1978} and the role of
screening and scattering. For the current configuration
shown in Fig. 1(b) the prediction of a simple Drude model, which
adds the conductivities of the two valleys with an effective mass
anisotropy ratio of $r=m_l/m_t=5.12$ and assumes a fixed scattering
time, is shown in Fig. 1(d) with a dashed line. This curve is only
adjusted to match the kink position and the resistance minimum of
the $n=3.93 \times 10^{11}$ cm$^{-2}$ data. This model predicts the
ratio of the resistances from saturation to balance to be
$R_e^{[110]}=(r+1)^2/4r=1.83$. We point out that there might be additional
contributions from screening and scattering effects. Screening
becomes less effective with increasing valley polarization and can
cause an extra increase in the resistance at larger valley
polarizations \cite{Screening,GunawanNature}. In contrast to
screening, the inter-valley scattering is more pronounced when two
valleys are occupied and results in a larger resistance around
balance. Since the experimentally measured values of $R_e^{[110]}$ range
from 1.8 to 2.1 and the data show a faster rise in resistance with
strain compared to the simple Drude model, we conclude that the
piezoresistance at $B=0$ mainly comes from the mass anisotropy and
the loss of screening.

\begin{figure}
\centering
\includegraphics[scale=0.86]{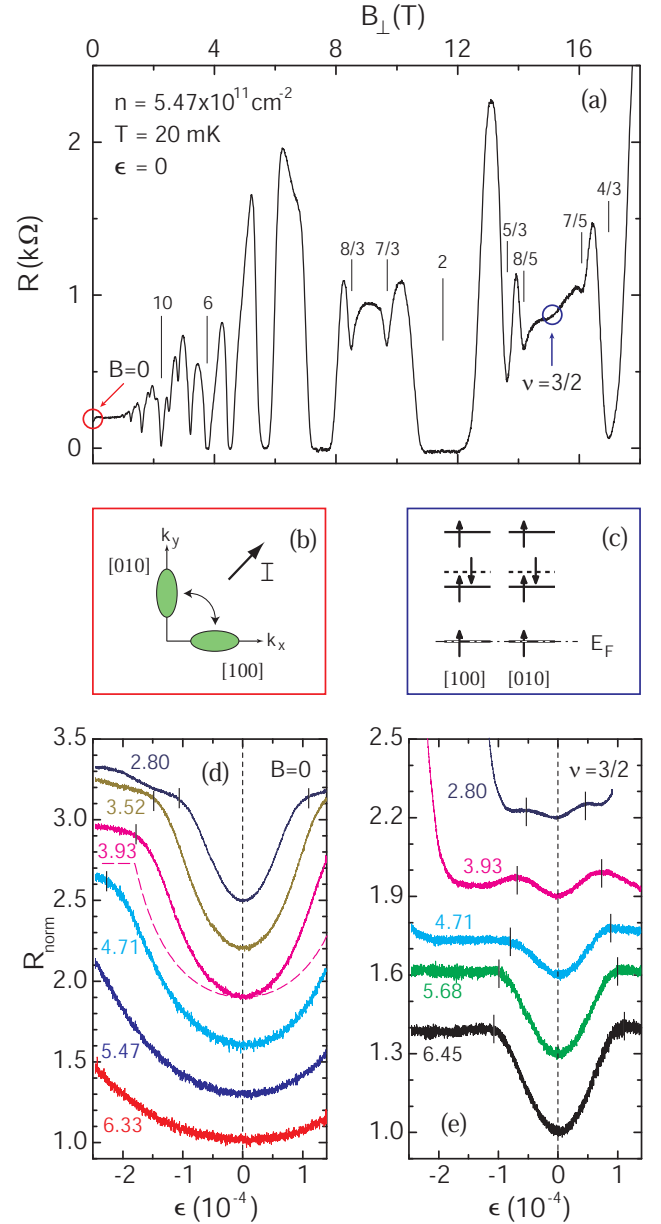}
\caption{(Color online) (a) Magnetoresistance ($R$ vs. $B_{\perp}$)
trace at zero strain (balanced valleys). (b) Schematic showing the
Fermi contours of the electrons in two valleys and the strain
induced inter-valley electron transfer. The current ($I$) is applied
along [110]. (c) Energy level diagram at $\nu=3/2$ for balanced
valleys. (d) and (e) Piezoresistance traces at different densities
for electrons and CFs, respectively. The values of density are given
in units of $10^{11}$ cm$^{-2}$, and the positions of the "kinks"
are marked by vertical lines.}
\end{figure}

The energy level diagram at $\nu=3/2$ for balanced valleys and
piezoresistance of CFs are shown in Figs. 1(c) and (e),
respectively. At high densities, the piezoresistance of CFs exhibits
features qualitatively similar to the piezoresistance of electrons:
the resistance increases as strain is swept away from zero and then
saturates at high strain values, signaling the full valley
polarization of CFs. However, there are several differences. The
resistance ratio $R_{CF}^{[110]}$ for CF piezoresistance from balance to
saturation is smaller than $R_e^{[110]}$ and the kink occurs at smaller
strain values for CFs compared to electrons. Furthermore, at low
densities, we observe a dramatic rise in the piezoresistance at
higher strains beyond the kink. We address these points in the
following three paragraphs.

\begin{figure*}
\centering
\includegraphics[scale=0.9]{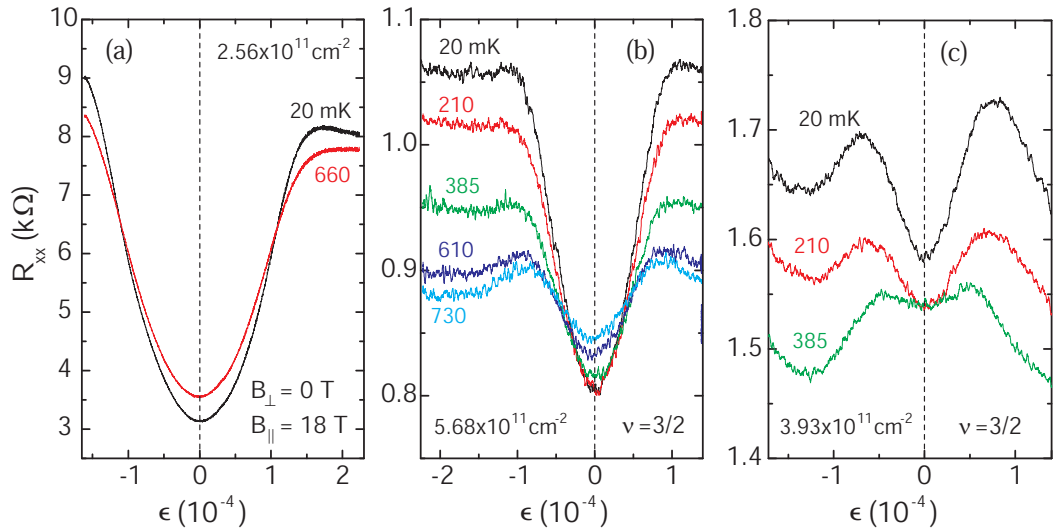}
\caption{(Color online) (a) Piezoresistance traces at different
temperatures for spin polarized electrons. (b) and (c)
Piezoresistance traces at different temperatures for CFs at
$\nu=3/2$ for two different densities.}
\end{figure*}

By making an analogy to electrons, the kink position in the
$\nu=3/2$ piezoresistance traces can be associated with the full
valley polarization of CFs, and the strain value at the kink
position gives the valley splitting energy that is equal to the
Fermi energy of CFs. Since the CF Fermi sea is a direct
manifestation of the Coulomb interaction, the valley splitting
energy needed to valley polarize the CFs is determined by the
Coulomb energy which is quantified by the magnetic length
\cite{BishopPRL07,PadmanabhanPRB09}. The kink positions at $\nu=3/2$
piezoresistance traces in Fig. 1(e) are indeed in agreement with the
results of previous studies of CF valley polarization energies,
determined from coincidence measurements of CF Landau levels
\cite{BishopPRL07,PadmanabhanPRB09}.

%The
%interpretation of $R_{CF}^{[110]}$ is more challenging than $R_e^{[110]}$.

The experimentally measured value of the resistance ratio $R_{CF}^{[110]}$
for CFs from balance to saturation is 1.4 for the highest density,
and drops almost to unity for the lowest densities. We note that the experimentally measured value of $R_{CF}^{[110]}$ is smaller than $R_e^{[110]}$. Similar to $R_e^{[110]}$, we expect that $R_{CF}^{[110]}$ would be affected by the effective
mass anisotropy of CFs and the
screening/scattering effects. Our piezoresistance resistance measurements along the [100] direction indeed show that CFs inherit the transport anisotropy of electrons at zero field, suggesting an anisotropy in CF effective mass \cite{GokmenUnpublished}. However, we emphasize that the observed transport anisotropy of CFs along the [100] is smaller compared to electrons and therefore qualitatively in agreement with the observation of $R_{CF}^{[110]} < R_e^{[110]}$.

%However, the application of a simple
%Drude model to the case CFs is not possible, because the effect of
%the effective mass anisotropy of electrons on CFs is unclear at the
%moment \cite{BalagurovPRB2000}. Moreover, as we discuss later,
%$R_{CF}^{[110]}$ strongly depends on temperature. Because of these
%complication it is not possible to distinguish the separate
%contributions of the mass anisotropy and screening/scattering to
%piezoresistance of CFs.

Another feature of the piezoresistance traces at $\nu=3/2$ for low
densities is the significant increase in resistance for high
strains. The reason for this increase is the coincidence of the
\textit{electron} Landau levels. As can be surmised from Fig. 1(c),
for sufficiently large strains, when the valley splitting energy is
equal to the electron cyclotron energy, the lowest electron Landau
level of one valley coincides with the second Landau level of the
other valley. Beyond this coincidence, the electrons are fully
valley polarized. Note that the increase in resistance at $\nu=3/2$
occurs roughly at the same strain value where the $B=0$
piezoresistance traces show the kink and signal full electron valley
polarization. We add that at much higher strains, well past the
coincidence, the piezoresistance at $\nu=3/2$ saturates again,
consistent with the $B=0$ data.

Now we present the temperature dependence of the CF piezoresistance
at $\nu=3/2$. In Figs. 2(b) and 2(c) we show piezoresistance traces
for CFs at two densities. The data in Fig. 2(b) reveal that, at high
densities and for small strains so that the valley polarization is
small, CFs exhibit a metallic behavior ($dR/dT > 0$). With
increasing strain, however, the resistance turns insulating ($dR/dT < 0$) as CFs become valley polarized. This observation demonstrates the importance of the discrete degrees of freedom for the temperature dependence of resistance of
CFs. At lower densities (Fig. 2(c)) the metallic behavior around
zero strain disappears, and the CFs act insulating in the full
strain range, including at $\epsilon=0$ where they are valley
degenerate.

Before presenting more details of our data, we briefly discuss the
temperature dependence of the resistance of 2D \textit{electrons}
and the related MIT. The scaling theory predicts an insulating phase
for a non-interacting 2DES with arbitrarily weak disorder, thanks to
the weak localization of electrons \cite{AbrahamsPRL1979}. However,
experiments in high quality Si 2DESs \cite{KravchenkoPRB1994} showed
that, at high densities, the 2DES exhibits a metallic temperature
dependence, and that the system can be driven to an insulating phase
by lowering the 2DES density (increasing the disorder). Similar
behavior has been reported for several other 2DESs
\cite{AbrahamsRevModPhys01}. Although there is no consensus whether or not
there exists a true metallic ground state (in the limit of zero
temperature) for a low-disorder 2DES, it is generally believed that
the observed metallic behavior is due to the interplay of disorder,
interaction and finite temperature effects. Additional measurements
have also shown that the spin polarization of the 2DES plays a
critical role in the apparent MIT: The system's behavior changes from metallic to insulating with increasing spin polarization \cite{Screening}.
A recent study \cite{GunawanNature} revealed that, for AlAs 2DESs,
not only spin but also valley polarization is an important parameter
for the apparent MIT, namely, the 2DES exhibits an insulating
behavior when both valley and spin polarizations pass beyond some
threshold value.

We illustrate this behavior for our AlAs 2DES sample in Fig. 2(a)
where we show piezoresistance traces for electrons at a very high
parallel magnetic field such that the 2D electrons are fully spin
polarized ($B_{\perp}=0$ in Fig. 2(a) trace). Both qualitatively and
quantitatively our results are consistent with the results of Ref.
\cite{GunawanNature}. At zero and small parallel fields, where the
2DES is not spin-polarized, it shows a metallic behavior in the
entire strain range (data not shown). But when the 2DES is fully
spin polarized via the application of a large parallel magnetic
field, increasing the valley polarization changes the temperature dependence of resistance from metallic to insulating (Fig. 2(a)).

Our $\nu=3/2$ data in Fig. 2(b) show that the CFs qualitatively
behave like the electrons. Note that because of the large g-factor,
the Zeeman energy in our sample is comparable to and even larger
than the cyclotron energy, as illustrated by the energy level
diagram in Fig. 1(c). Therefore, the CFs at $\nu=3/2$ are fully spin
polarized while we control their valley occupation via the
application of strain. The electron data of Fig. 2(a) and CF data of
Fig. 2(b) imply that, when the spin degree of freedom is frozen,
(high density) electrons and CFs both show a metallic behavior when
they have a valley degree of freedom. However, both systems show insulating behavior when the valley
polarization is above some threshold value.

We observe the metallic behavior in the absence of strain and the
valley polarization driven MIT for CFs in the high density range
($n>3.93 \times 10^{11}$ cm$^{-2}$). As the density is lowered, the
metallic temperature dependence disappears and CFs exhibit an
insulating behavior regardless of their valley polarization (Fig.
2(c)). This is consistent with the increasing disorder in the system
at lower densities, and similar to the disorder-induced MIT of CFs demonstrated by Ref. \cite{LiangSSC1997}.
We emphasize, however, that the CFs seem to be more sensitive to
disorder than electrons: In the absence of strain, the density below
which the \textit{electrons} exhibit an insulating behavior is $\sim 1
\times 10^{11}$ cm$^{-2}$ while for the $\nu=3/2$ CFs it is $\sim 4
\times 10^{11}$ cm$^{-2}$.

Our piezoresistance results demonstrate that CFs and electrons show
qualitatively very similar behaviors. However, they differ
quantitatively in several ways. First, the valley splitting energy
required to completely valley polarize the CFs is smaller than the
electrons and this difference is reasonably well understood
\cite{BishopPRL07,PadmanabhanPRB09}. Second, the piezoresistance
ratio $R_{CF}^{[110]}$ from balance to saturation for CFs is smaller than
the corresponding ratio $R_e^{[110]}$ for electrons and has a stronger
temperature dependence. Third, the temperature dependence of the
resistance, and the observation of an insulating behavior for CFs with
increasing valley polarization or decreasing density, are
qualitatively similar to electrons.

We thank the NSF for support. Part of this work was performed at the
NHMFL, Tallahassee, which is also supported by the NSF. We thank E.
Palm, T. Murphy, J. Park, and G. Jones for assistance.

\break

\end{document}